\documentclass{article}
\usepackage{lmodern}

\usepackage[T1]{fontenc}
\usepackage[latin9]{inputenc}
\usepackage{geometry}
\usepackage[english]{babel}
\usepackage{array}
\usepackage{longtable}
\usepackage{float}
\usepackage{amsmath}
\usepackage{amsthm}
\usepackage{amssymb}
\usepackage{graphicx}
\usepackage[round]{natbib}
\usepackage[pdfusetitle,
 bookmarks=true,bookmarksnumbered=false,bookmarksopen=false,
 breaklinks=false,pdfborder={0 0 0},pdfborderstyle={},backref=false,colorlinks=false]
 {hyperref}

\makeatletter

\providecommand{\tabularnewline}{\\}

\usepackage{algcompatible}
\usepackage{algorithm}

\makeatother

\begin{document}
\title{Enhancing Fourier pricing with machine learning}
\author{Gero Junike\thanks{Corresponding author. Carl von Ossietzky Universität, Institut für
Mathematik, 26129 Oldenburg, Germany, ORCID: 0000-0001-8686-2661,
Phone: +49 441 798-3729, E-mail: gero.junike@uol.de}, Hauke Stier\thanks{Carl von Ossietzky Universität, Institut für Mathematik, 26129 Oldenburg,
Germany.}}
\maketitle
\begin{abstract}
Fourier pricing methods such as the Carr-Madan formula or the COS
method are classic tools for pricing European options for advanced
models such as the Heston model. These methods require tuning parameters
such as a damping factor, a truncation range, a number of terms, etc.
Estimating these tuning parameters is difficult or computationally
expensive. Recently, machine learning techniques have been proposed
for fast pricing: they are able to learn the functional relationship
between the parameters of the Heston model and the option price. However,
machine learning techniques suffer from error control and require
retraining for different error tolerances. In this research, we propose
to learn the tuning parameters of the Fourier methods (instead of
the prices) using machine learning techniques. As a result, we obtain
very fast algorithms with full error control: Our approach works with
any error tolerance without retraining, as demonstrated in numerical
experiments using the Heston model.
\end{abstract}

\section{Introduction}

Fourier methods, such as the Carr-Madan formula and the COS method,
see \citet{carr1999option} and \citet{fang2009novel}, are widely
used to price European options. In order to speed up option pricing,
\citet{liu2019neural,liu2019pricing}, \citet{yang2017gated} and
\citet{sirignano2018dgm} propose a prediction of option prices using
neural networks. \citet{ruf2019neural} provide a comprehensive review
of neural networks for option pricing. \citet{liu2019neural,liu2019pricing}
use a parametric approach and consider an advanced stock price model,
such as the Heston model, see \citet{heston1993closed}. They use
a set of market parameters, including strike price and maturity, as
well as model parameters, to predict the corresponding option prices.
\citet{de2018machine} use machine learning techniques based on Gaussian
process regression for prediction of option prices. 

While \citet{de2018machine} and \citet{liu2019neural,liu2019pricing}
were able to accelerate the existing Fourier methods to some extent,
their approaches also exhibited certain limitations. \citet{liu2019neural,liu2019pricing}
obtain a mean absolute error (MAE) of about $10^{-4}$. \citet{de2018machine}
also obtains a MAE of about $10^{-4}$ and a maximum absolute error
of approximately $10^{-3}$ on their sample. \citet[Table 2]{de2018machine}
compare the numerical effort with the Carr-Madan formula and obtain
an acceleration factor between 10 and 40 for European options.

However, the approaches described in \citet{liu2019neural,liu2019pricing}
and \citet{de2018machine} suffer from a lack of error control: To
achieve higher numerical pricing accuracy, deeper neural networks
are necessary and the machine learning methods need to be retrained
with more samples, which is very time-consuming and impractical in
most situations. 

In this paper, we propose an indirect use of machine learning methods
to improve the accuracy and efficiency of existing pricing techniques
with full error control. We focus on the COS method, but our approach
is also applicable to other methods, i.e., we also discuss the Carr-Madan
formula.

We describe the main idea of the COS method, details can be found,
e.g., in \citet{fang2009novel,oosterlee2019mathematical,junike2022precise}:
Given only the characteristic function of the log-returns of the underlying,
the density of the log-returns is approximated in two steps: i) truncate
the density on a finite interval $[a,b]$ and ii) approximate the
truncated density by a finite Fourier-cosine approximation with $N$
terms. There is a clever trick to obtain the cosine-coefficients of
the truncated density efficiently from the characteristic function.
The CPU time of the COS method depends linearly on the number of terms
$N$. Note that the choice of the truncation range has a significant
influence on the number of terms required to achieve a certain accuracy.
There are explicit formulas for the truncation range and the number
of terms depending on an error tolerance $\varepsilon>0$, see \citet{junike2022precise}
and \citet{junike2023handle}. However, the truncation range formula
requires evaluating higher-order derivatives of the characteristic
function, which can be very time-consuming, e.g., in the case of the
Heston model. The formula for the number of terms requires integration
of the product of the characteristic function and a polynomial, which
is also very time consuming. Fortunately, the time-consuming part
required to obtain $[a,b]$ and $N$ \emph{does not depend on the
required error tolerance $\varepsilon$}.

In this paper, we use machine learning techniques to learn the $n$-th
derivatives of the characteristic function evaluated at zero and learn
the integral of the characteristic function times a polynomial, which
is independent of the required error tolerance. Then, we use these
predicted values and the error tolerance to obtain the truncation
range and the number of terms. The COS method can then be applied
to price European options.

Different traders may use different error tolerances, but our machine
learning techniques do not require retraining. This error control
is an advantage over direct prediction of option prices by machine
learning techniques. The actual calculation of the option price using
the COS method is then very fast.

The paper is structured as follows. Section \ref{sec:Heston-model}
gives an overview of the Heston model, which will be used in the numerical
experiments. In Section \ref{sec:Methodology}, we introduce the COS
method and the Carr-Madan formula and machine learning techniques.
Section \ref{sec:Numerical-experiments} provides the numerical experiments
to demonstrate the performance of the proposed method. Section \ref{sec:Conclusion}
concludes the paper.

\section{The Heston model\protect\label{sec:Heston-model}}

Consider a financial market with a riskless bank-account and a stock
with deterministic price $S_{0}>0$ today and random price $S_{T}$
at some future date $T>0$. In the Heston model with parameters $\kappa>0$,
$\theta>0$, $\xi>0$, $\rho\in[-1,1]$ and $v_{0}>0$, the stock
price is described by the following system of differential equations

\begin{eqnarray}
\frac{dS_{t}}{S_{t}} & = & rdt+\sqrt{v_{t}}dW_{t},\,\,\,S_{0}\geq0\label{eq:heston}\\
dv_{t} & = & \kappa(\theta-v_{t})dt+\xi\sqrt{v_{t}}dZ_{t}.\label{eq:CIR}
\end{eqnarray}
$W$ and $Z$ are correlated Brownian motions such that $\text{cov}\left[dW_{t}dZ_{t}\right]=\rho dt$,
see \citet{heston1993closed}. 

The CIR process, described by Equation (\ref{eq:CIR}), stays positive
if $2\kappa\theta\geq\xi^{2}$, which is known as the \emph{Feller
condition}, see \citet{andersen2007moment}. The characteristic function
of the log stock price, see \citet{bakshi1997empirical}, is given
by

\begin{eqnarray*}
\varphi_{\log\left(S_{t}\right)}(u) & = & E\left[\exp(iu\log\left(S_{t}\right)\right]\\
 & = & \exp\left(iu\left(\log S_{0}+rt\right)\right)\\
 &  & \times\exp(\theta\kappa\xi^{-2}\left(\left(\kappa-\rho\xi ui-d\right)t-2\log\left(\frac{1-ge^{-dt}}{1-g}\right)\right))\\
 &  & \times\exp\left(v_{0}\xi^{-2}\frac{\left(\kappa-\rho\xi iu-d\right)\left(1-e^{-dt}\right)}{1-ge^{-dt}}\right),
\end{eqnarray*}
where
\begin{eqnarray*}
d & = & \left(\left(\rho\xi ui-\kappa\right)^{2}-\xi^{2}\left(-iu-u^{2}\right)\right)^{\frac{1}{2}},\\
g & = & \frac{\kappa-\rho\xi ui-d}{\kappa-\rho\xi ui+d}.
\end{eqnarray*}

\section{Algorithms: Numerical tools and machine learning\protect\label{sec:Methodology}}

\subsection{The Carr-Madan formula}

\citet{carr1999option} showed that the price of a European call option
with strike $K$ and time to maturity $T$ is given by 
\begin{equation}
e^{-\alpha\log(K)}e^{-rT}\frac{1}{\pi}\int_{0}^{\infty}\Re\left\{ e^{-iv\log(K)}\frac{\varphi_{\log(S_{T})}\big(v-i(\alpha+1)\big)}{\alpha^{2}+\alpha-v^{2}+i(2\alpha+1)v}\right\} dv,\label{eq:carr-madan=000020formula}
\end{equation}
where $\alpha>0$ is a damping factor such that $E[S_{T}^{1+\alpha}]<\infty$
and $\varphi_{\log(S_{T})}$ is the characteristic function of $\log(S_{T})$.
$\Re(z)$ denotes the real part of a complex number $z$ and $i=\sqrt{-1}$
is the complex unit. The integral in Eq. (\ref{eq:carr-madan=000020formula})
can be truncated to $(0,M)$, for some $M>0$, and then be evaluated
using, e.g., Simpson's rule with $N$ grid points.

\subsection{The COS method}

We summarize the COS method. This section is based on \citet{fang2009novel},
\citet{junike2022precise} and \citet{junike2023handle}. Let $\mu$
be the expectation of $\log(S_{T})$ under the risk-neutral measure
and assume that the characteristic function $\varphi_{X}$ of the
centralized log-returns $X:=\log(S_{T})-\mu$ is given in closed-form.
The function $\varphi_{X}$ is explicitly given for many models such
as the Heston model. The price of a European put option with maturity
$T>0$ and strike $K>0$ is given by
\begin{equation}
\int_{\mathbb{R}}e^{-rT}\max\left(K-e^{x+\mu},0\right)f(x)dx,\label{eq:price}
\end{equation}
where $f$ is the density of $X$. The price of a call option can
be obtained by the put-call-parity. Very often, $f$ is not explicitly
given and the COS method can be used to approximate $f$ and the price
of the option.

For some $L>0$, the density $f$ is truncated and the truncated density
is approximated by a cosine series expansion:
\begin{equation}
f(x)\approx\frac{c_{0}}{2}+\sum_{k=1}^{N}c_{k}\cos\left(k\pi\frac{x+L}{2L}\right),\quad x\in[-L,L],\label{eq:f}
\end{equation}
where for $k=0,1,...,N$, the coefficients $c_{k}$ are defined by
\begin{align}
c_{k}:=\frac{1}{L}\int_{\mathbb{R}}f(x)\cos\big(k\pi\frac{x+L}{2L}\big)dx= & \frac{1}{L}\Re\bigg\{\varphi\bigg(\frac{k\pi}{2L}\bigg)e^{i\frac{k\pi}{2}}\bigg\}.\label{eq:ck}
\end{align}
The second Equality in (\ref{eq:ck}) follows from a simple analysis,
see \citet{fang2009novel}. The price of a European put option can
be approximated by replacing $f$ in (\ref{eq:price}) with its approximation
(\ref{eq:f}), which gives
\[
\int_{\mathbb{R}}e^{-rT}\max\left(K-e^{x+\mu},0\right)f(x)dx\approx\frac{c_{0}v_{0}}{2}+\sum_{k=1}^{N}c_{k}v_{k},
\]
where
\begin{align*}
v_{k} & :=\int_{-L}^{L}e^{-rT}\max\left(K-e^{x+\mu},0\right)\cos\big(k\pi\frac{x+L}{2L}\big)dx,\quad k\in\{0,1,2,...\}.
\end{align*}
The coefficients $c_{k}$ are given in closed form when $\varphi_{X}$
is given analytically and the coefficients $v_{k}$ can also be computed
explicitly in important cases, e.g., for plain vanilla European put
or call options and digital options, see \citet{fang2009novel}. This
makes the COS method numerically very efficient and robust. 

We provide formulas for the coefficients $v_{k}$ for a European put
option: Let $d:=\min\left(\log\left(K\right)-\mu,L\right)$. For a
European put option, it holds that $v_{k}=0$ if $d\leq-L$ and otherwise
\[
v_{k}=e^{-rT}\left(K\Psi_{0}(k)-e^{\mu}\Psi_{1}(k)\right),
\]
where

\[
\Psi_{0}(k)=\begin{cases}
d+L & ,k=0\\
\frac{2L}{k\pi}\sin\big(k\pi\dfrac{d+L}{2L}\big) & ,k>0
\end{cases}
\]
and
\begin{align*}
\Psi_{1}(k) & =\frac{e^{d}\left(\frac{k\pi}{2L}\sin\left(k\pi\frac{d+L}{2L}\right)+\cos\left(k\pi\frac{d+L}{2L}\right)\right)-e^{-L}}{1+\left(\frac{k\pi}{2L}\right)^{2}},\quad k\geq0,
\end{align*}
see \citet[Appendix A]{junike2022precise}. To price a call option
it is numerically more stable to price a put option instead and use
the put-call parity, see \citet{fang2009novel}.

To apply the COS method, one has to specify the truncation range $[-L,L]$
and the number of terms $N$. For a given error tolerance $\varepsilon$
small enough, both parameters can be chosen as follows to ensure an
approximation error smaller than $\varepsilon$, see \citet{junike2022precise}
and \citet{junike2023handle}. If $\varepsilon$ is small enough and
$f$ has semi-heavy tails, the truncation range of a put option can
be chosen using Markov's inequality by
\begin{equation}
L=L(\varepsilon,\mu_{n})=\mu_{n}\times\left(\frac{2Ke^{-rT}}{\varepsilon}\right)^{\frac{1}{n}},\label{eq:COS-=000020truncation=000020range}
\end{equation}
where $n\in\mathbb{N}$ is even and $\mu_{n}$ is the $n$-th root
of the $n$-th moment of $X$, which can be obtained using a computer
algebra system and the relation 
\begin{equation}
\mu_{n}=\sqrt[n]{\frac{1}{i^{n}}\left.\frac{\partial^{n}}{\partial u^{n}}\varphi_{X}(u)\right|_{u=0}}.\label{eq:moment=000020via=000020derivative}
\end{equation}
Often, $n\in\{4,6,8\}$ is a reasonable choice, see \citet[Cor. 9]{junike2022precise}.
If $f$ is also $s+1\in\mathbb{N}$ times differentiable with bounded
derivatives, then the number of terms can be chosen by
\begin{equation}
N=N(\varepsilon,I_{s})=I_{s}\times\left(\frac{2^{s+\frac{5}{2}}L^{s+2}}{s\pi^{s+1}}\frac{12Ke^{-rT}}{\varepsilon}\right)^{\frac{1}{s}},\label{eq:COS-number=000020of=000020terms}
\end{equation}
where 
\begin{equation}
I_{s}:=\left(\frac{1}{2\pi}\int_{\mathbb{R}}|u|^{s+1}|\varphi_{X}(u)|du\right)^{\frac{1}{s}},\label{eq:=000020supremum=000020norm}
\end{equation}
see \citet[Eq. (3.8)]{junike2023handle}. The last integral can be
solved numerically by standard techniques and in some cases it is
given explicitly. One should choose $s$ such that the left-hand side
of Inequality (\ref{eq:COS-number=000020of=000020terms}) is minimized.
For the Heston model, $s$ is set to $s=20$ in \citet{junike2023handle}.
An implementation of the truncation range, the number of terms and
the COS method for the Heston model can found in Appendix \ref{subsec:=000020code}.

\subsection{Machine learning techniques\protect\label{subsec:Machine-Learning-models}}

\paragraph{Decision Tree:}

\begin{figure}[H]
\begin{centering}
\includegraphics[scale=0.5]{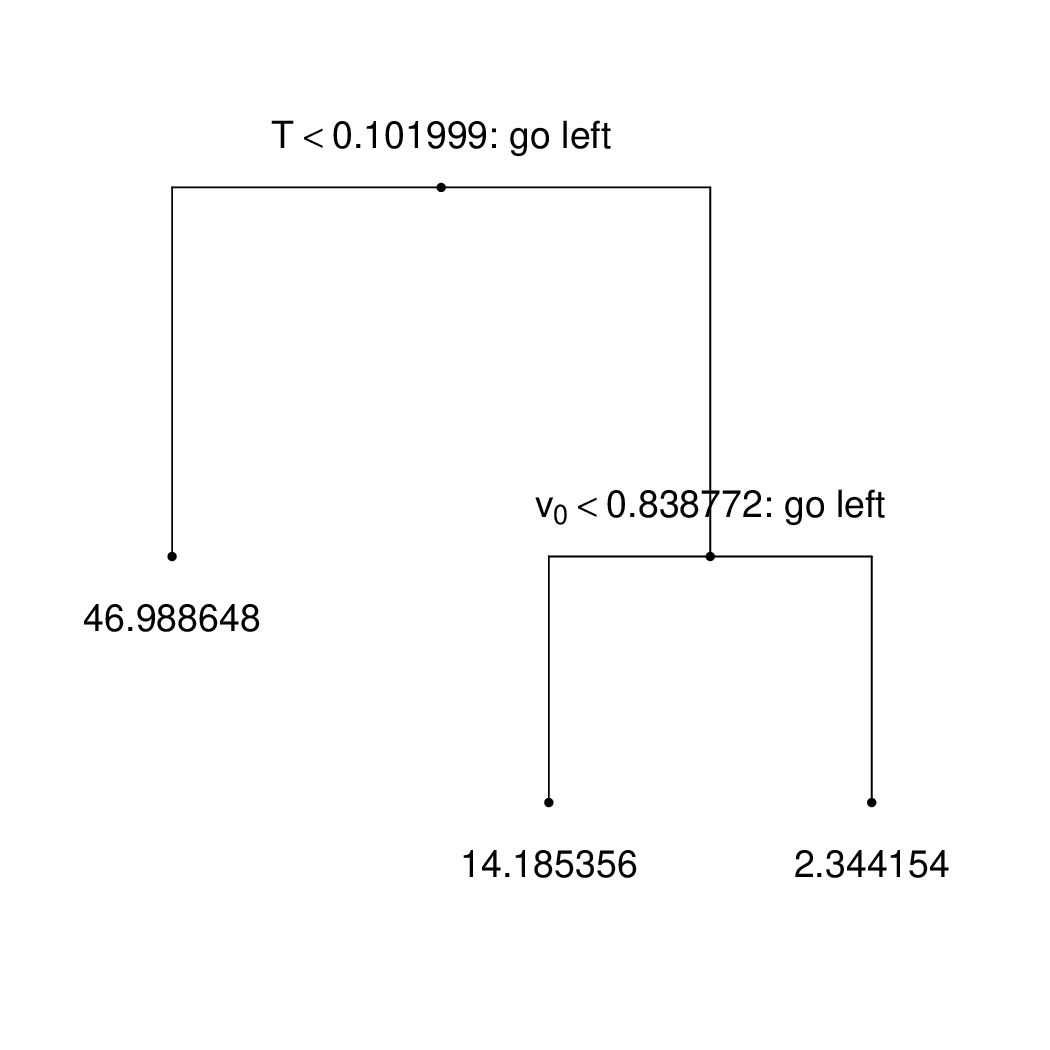}
\par\end{centering}
\caption{\protect\label{fig:Example=000020Decision=000020tree}Example of a
DT as in Table \ref{tab:Example=000020Decision=000020tree}. }
\end{figure}

Decision trees (DT), see \citet{breiman1984classification}, operate
by recursively partitioning the input data into subsets, thereby forming
a tree-like structure, see Table \ref{tab:Example=000020Decision=000020tree}
and Figure \ref{fig:Example=000020Decision=000020tree}. At each internal
node of the DT, the algorithm selects a feature and a threshold value
to split the data into two subsets. 

For example, in the first row of Table \ref{tab:Example=000020Decision=000020tree},
all input values with maturity $T$ less than or equal to $0.1019998$
are assigned to node $1$, all other values are assigned to node $2$.
The goal of these splits is to create child nodes with greater homogeneity.
The recursive splitting process continues until a stopping criterion
is met, such as a maximum tree depth or a minimum node size for splitting.

To build a DT for regression, the splitting is based on variance reduction.
The algorithm selects the features and thresholds that most strongly
reduce the variance at each node for splitting.

Given new samples, predictions are made at the leaf nodes, where the
model assigns the average of the data points within the node. This
simplicity and transparency make DT highly effective at handling complex
data sets while maintaining interpretability.

\begin{table}[H]
\begin{centering}
\begin{tabular}{|c|c|c|c|>{\centering}p{2.5cm}|>{\centering}p{2.5cm}|c|}
\hline 
nodeID & leaf node & variable & split value & left-child\\
 (if variable $<$ split value) & right-child \\
(if variable $\geq$ split value) & prediction\tabularnewline
\hline 
\hline 
0 & No & $T$ & 0.101999 & 1 & 2 & NA\tabularnewline
\hline 
1 & Yes & NA & NA & NA & NA & 46.988648\tabularnewline
\hline 
2 & No & $v_{0}$ & 0.838772 & 3 & 4 & NA\tabularnewline
\hline 
3 & Yes & NA & NA & NA & NA & 14.185356\tabularnewline
\hline 
4 & Yes & NA & NA & NA & NA & 2.344154\tabularnewline
\hline 
\end{tabular}\caption{\protect\label{tab:Example=000020Decision=000020tree}Example of a
DT. }
\par\end{centering}
\end{table}

\paragraph{Random Forest:}

Random forests (RF), see \citet{breiman2001random} are an ensemble
of DTs to improve the accuracy and robustness of predictions. Each
DT in the RF is trained on a random subset of the data using bootstrap
aggregation. At each node, a random subset of the features is used
for the splitting. In a RF, each DT makes a prediction independently
and the final output is determined by averaging the individual predictions
of each single tree.

\paragraph{Neural Networks:}

A neural network (NN) consists of one or more layers, each consisting
of a number of artificial neurons, see \citet{goodfellow2016deep}.
A single neuron transforms its multidimensional input $x\in\mathbb{R}^{n}$
into a one-dimensional output. For some weights $w\in\mathbb{R}^{n+1}$,
the weighted mean of the input is then transformed by an activation
function $g:\mathbb{R}\to\mathbb{R}$, i.e., the output of a neuron
is given by $g\left(\sum_{i=1}^{n}w_{i}x_{i}+w_{n+1}\right).$ Examples
of activation functions are the ReLU function $g(y)=\max(y,0)$ or
the Sigmoid function $g(y)=\frac{1}{1+e^{-x}}$. In the first layer
of the NN, the neurons receive the input data and the output of each
neuron is passed to all neurons in the following layers until the
last layer is reached.

At the start of training, the weights of the NN are randomly initialized.
During the training phase, the weights are chosen in such a way that
the functional relationship between input and output data is mapped
as well as possible. 

In this work, we test the following regularization techniques that
can improve the robustness of the NN: Dropout means randomly deactivating
some neurons. Gaussian noise is a regularization technique that adds
normally distributed numbers with zero mean and small variance to
each weight at each update step. Batch normalization standardizes
the inputs of each layer. These and other regularization techniques
are discussed in detail in, for example, \citet{goodfellow2016deep}.

\section{Numerical experiments\protect\label{sec:Numerical-experiments}}

In this section, we use the machine learning techniques DT, NN and
RF to predict the tuning parameters of the Carr-Madan formula and
the COS method. For training, we randomly generate parameters of the
Heston model. The ranges of the six parameters are shown in Table
\ref{tab:Range-of-parameters}. The wide ranges of these parameters
include parameters that are typically used for the Heston model, see
\citet{andersen2007efficient,crisostomo2015analysis,cui2017full,engelmann2021calibration,fang2009novel,forde2012small,levendorskiui2012efficient}
and \citet{schoutens2003perfect}.

For each sample (consisting of the five parameters for the Heston
model and the maturity), we compute $\mu_{4}$ and $\mu_{8}$ and
$I_{20}$ for the entire data set, using Eqs. (\ref{eq:moment=000020via=000020derivative},
\ref{eq:=000020supremum=000020norm}). The derivatives of $\varphi_{X}$
are calculated using a computer algebra system. As a side note: One
may also approximate the moments as in \citet{choudhury1996numerical}
to avoid the computation of the derivatives.

We exclude all the model parameters for which Eq. (\ref{eq:moment=000020via=000020derivative})
gives negative results, assuming that the moments do not exist in
these cases and we remove all parameters for which the Feller condition
$2\kappa\eta\geq\xi^{2}$ is not satisfied.

In the following numerical experiments, we price a European call option
with $S_{0}=100$, strike $K=100$ and interest rate $r=0$. We also
tested other strikes, i.e., $K\in\{75,125\}$ and obtained similar
results. For each sample, we calculate a reference price. To obtain
the reference prices we use the COS method with truncation range $L(\varepsilon,\mu_{8})$
and number of terms $N(\varepsilon,I_{20})$, where we set $\varepsilon=10^{-9}$.
To confirm the prices we use the Carr-Madan formula with truncation
range $M=1024$, $N=2^{20}$ and appropriate damping factors. We remove
a few samples where the prices were too unstable and the COS method
and the Carr-Madan formula give completely different results. For
all remaining options, the COS method and the Carr-Madan formula coincide
at least up to seven decimal place.

We receive a cleaned data set of $250,000$ samples. We take $100,000$
samples for training and validation and use the remaining $50,000$
samples as a test set. All experiments are run on a laptop with an
Intel i7-11850H processor and 32 GB of RAM.

\begin{table}[H]
\begin{centering}
\begin{tabular}{|c|c|}
\hline 
Parameter & Value range\tabularnewline
\hline 
\hline 
speed of mean reversion $\kappa$ & $[10^{-3},10]$\tabularnewline
\hline 
level of mean reversion $\theta$ & $[10^{-3},2]$\tabularnewline
\hline 
volatility of variance $\xi$ & $[10^{-2},5]$\tabularnewline
\hline 
correlation coefficient $\rho$ & $[-0.99,0.99]$\tabularnewline
\hline 
initial variance $v_{0}$ & $[10^{-3},2]$\tabularnewline
\hline 
time to maturity $T$ & $[250^{-1},10]$\tabularnewline
\hline 
\end{tabular}
\par\end{centering}
\caption{\protect\label{tab:Range-of-parameters}Range of parameters of the
Heston model, including parameters that are typically used.}

\end{table}

\subsection{On the tuning parameters of the COS method}

To apply the COS method, we use the formulas for the truncation range
and the number of terms in Eq. (\ref{eq:COS-=000020truncation=000020range})
and (\ref{eq:COS-number=000020of=000020terms}). For the Heston model,
it is time-consuming to compute $\mu_{8}$ in Eq. (\ref{eq:moment=000020via=000020derivative})
and to solve the integral $I_{20}$ in Eq. (\ref{eq:=000020supremum=000020norm}).
Therefore, we use the machine learning techniques DT, RF and NN for
a fast estimation of $\mu_{8}$ and $I_{20}$. 

To identify an appropriate architecture for the different machine
learning techniques, we perform a rough hyperparameter optimization.
For the DT, we optimize over the maximum depth and the minimum node
size. In addition, the number of DTs in the RF is optimized, resulting
in the hyperparameters shown in Table \ref{tab:hyperparameter=000020DT=000020RF}.
The R package \emph{ranger} is used for both DT and RF. We consider
a big DT (bDT) of arbitrary depth and a small DT (sDT) of depth 5.
The sDT for $\mu_{8}$ and the sDT for $I_{20}$ are tabulated in
Appendix \ref{subsec:=000020code} and \ref{subsec:=000020code} and
could be implemented directly without using additional software packages.

\begin{table}[H]
\begin{centering}
\begin{tabular}{|c|>{\centering}p{1cm}|>{\centering}p{1.5cm}|>{\centering}p{1cm}|>{\centering}p{1cm}|>{\centering}p{1cm}|>{\centering}p{1cm}|}
\hline 
Parameters & bDT for $I_{20}$ & bDT for $\mu_{8}$ & RF for $I_{20}$ & RF for $\mu_{8}$ & sDT for $I_{20}$ & sDT for $\mu_{8}$\tabularnewline
\hline 
\hline 
Number of DT & 1 & 1 & 500 & 600 & 1 & 1\tabularnewline
\hline 
Maximal tree depth & $30$ & unlimited & 50 & 90 & 5 & 5\tabularnewline
\hline 
Minimal node size to split at & $8$ & $6$ & 1 & 1 & 5 & 5\tabularnewline
\hline 
\end{tabular}
\par\end{centering}
\caption{\protect\label{tab:hyperparameter=000020DT=000020RF}Selected hyperparameters
of the DT and RF.}

\end{table}

The architectural specifications of the NN are described in Table
\ref{tab:=000020hyperparameters=000020NN}. The NN is trained with
100 epochs, a validation split of $0.2$ and the mean squared error
(MSE) 
\[
\text{MSE}(y,\tilde{y})=\frac{1}{n}\sum_{i=1}^{n}(y_{i}-\tilde{y}_{i})^{2},\quad y,\tilde{y}\in\mathbb{R}^{d},
\]
as the loss metric. For the starting values of the weights we use
the He initialization, see \citet{he2015delving}. For the NN, we
use tensorflow via the keras package called from R. 

Table \ref{tab:MSE} shows the MSE on the test set for the different
machine learning techniques. It can be observed that for $\mu_{8}$,
the NN has a smaller MSE than the RF, while the bDT has a comparatively
large MSE. With regard to $I_{20}$, the RF has the smallest MSE,
while the MSE of the NN and the bDT are about $40\%$ larger. The
sDT has a significantly larger MSE for both $\mu_{8}$ and $I_{20}$.

\begin{table}[H]
\begin{centering}
\begin{tabular}{|c|>{\centering}p{4cm}|c|c|}
\hline 
Parameters & Optimization range & NN for $I_{20}$ & NN for $\mu_{8}$\tabularnewline
\hline 
\hline 
Hidden layers & $\{1,\dots,4\}$ & 4 & 3\tabularnewline
\hline 
Neurons & $\{32,64,128,...,2048\}$ & 1024, 256, 256, 32 & 256, 128, 32\tabularnewline
\hline 
Activation function & ReLU, Leaky ReLU, Sigmoid, ELU, tanh & Sigmoid & Sigmoid\tabularnewline
\hline 
Dropout rate & $\{0,0.1,0.2,...,0.5\}$ & 0.2 & 0\tabularnewline
\hline 
Noise rate & $\{0.01,0.02,...,0.1\}$ & 0.07 & 0.02\tabularnewline
\hline 
Optimizer & Adam, SGD, RMSProp & Adam & Adam\tabularnewline
\hline 
Batch normalization & yes, no & no & no\tabularnewline
\hline 
Batch size & $\{128,256,512,1024\}$ & 512 & 256\tabularnewline
\hline 
\end{tabular}
\par\end{centering}
\caption{\protect\label{tab:=000020hyperparameters=000020NN}Selected hyperparameters
of the NN.}
\end{table}

\begin{table}[H]
\begin{centering}
\begin{tabular}{|c|>{\centering}p{2.5cm}|>{\centering}p{2.5cm}|>{\centering}p{2.5cm}|>{\centering}p{2.5cm}|}
\hline 
 & RF & NN & bDT & sDT\tabularnewline
\hline 
\hline 
$\mu_{8}$ & 0.0703 & 0.0058 & 0.2764 & 2.2390\tabularnewline
\hline 
$I_{20}$ & 33.6615 & 44.2353 & 49.0372 & 61.9859\tabularnewline
\hline 
\end{tabular}
\par\end{centering}
\caption{\protect\label{tab:MSE}MSE of the prediction of $\mu_{8}$ and $I_{20}$
for different ML techniques on the test set.}
\end{table}

Next, we calculate the price of the call option for different model
parameters. We use the COS method with $L(\varepsilon,\mu_{4})$ or
$L(\varepsilon,\mu_{8})$ and $N(\varepsilon,I_{20})$, where $\varepsilon\in\{10^{-1},...,10^{-7}\}$. 

The Table \ref{tab:Accuracy=000020COS} shows the percentage of samples
in the test set for which the required accuracy is achieved by obtaining
$\mu_{n}$ and $I_{s}$ directly from Eqs. (\ref{eq:moment=000020via=000020derivative},
\ref{eq:=000020supremum=000020norm}), which is very time-consuming,
or by estimating $\mu_{n}$ and $I_{s}$ via DTs, RF or a NN, which
is very fast. The direct way of obtaining $\mu_{8}$ and $I_{20}$
and the estimation by the RF result in $100\%$ accurate option prices
on the test set for all $\varepsilon$. The NN also achieves a high
accuracy of about $99.98\%$ for all $\varepsilon$. This result could
be further improved with a different NN architecture and additional
training. It can be observed that a single bDT is also able to estimate
$I_{20}$ and $\mu_{8}$ with sufficient accuracy to price the call
option with different error bounds for at least $99.96\%$ of the
samples. And even a simple technique like the sDT already achieves
an accuracy of at least $98\%$ on the test set. 

These very good results are a consequence of the fact that the formulas
in Eq. (\ref{eq:COS-=000020truncation=000020range}) and (\ref{eq:COS-number=000020of=000020terms})
are derived using many inequalities, thus overestimating the minimum
truncation range $L$ and the number of terms $N$ needed to accurately
price the option. Therefore, a rough estimate of $\mu_{8}$ and $I_{20}$
is sufficient for precise option pricing.

\begin{table}[H]
\begin{centering}
\begin{tabular}{|c|>{\centering}p{1.8cm}|>{\centering}p{1.8cm}|>{\centering}p{1.8cm}|>{\centering}p{1.8cm}|>{\centering}p{1.8cm}|>{\centering}p{1.8cm}|}
\hline 
$\varepsilon$ & Ana. calc. of $\mu_{4}$ and num. integration of $I_{20}$ & Ana. calc. of $\mu_{8}$ and num. integration of $I_{20}$ & $\mu_{8}$ and $I_{20}$ via RF & $\mu_{8}$ and $I_{20}$ via NN & $\mu_{8}$ and $I_{20}$ via bDT & $\mu_{8}$ and $I_{20}$ via sDT\tabularnewline
\hline 
\hline 
$10^{-1}$ & 0.999\% & 100\% & 100\% & 100\% & 100\% & 99.904\%\tabularnewline
\hline 
$10^{-2}$ & 0.999\% & 100\% & 100\% & 99.996\% & 99.998\% & 99.684\%\tabularnewline
\hline 
$10^{-3}$ & 100\% & 100\% & 100\% & 99.994\% & 99.998\% & 99.320\%\tabularnewline
\hline 
$10^{-4}$ & 100\% & 100\% & 100\% & 99.988\% & 99.986\% & 98.824\%\tabularnewline
\hline 
$10^{-5}$ & 100\% & 100\% & 100\% & 99.986\% & 99.976\% & 98.512\%\tabularnewline
\hline 
$10^{-6}$ & 100\% & 100\% & 100\% & 99.988\% & 99.970\% & 98.288\%\tabularnewline
\hline 
$10^{-7}$ & 100\% & 100\% & 100\% & 99.990\% & 99.968\% & 98.192\%\tabularnewline
\hline 
\end{tabular}
\par\end{centering}
\caption{\protect\label{tab:Accuracy=000020COS}Accuracy of the COS method
for different error tolerances $\varepsilon$ on the test set for
a call option with $S_{0}=100$ and $K=100$ with $\mu_{4}$, $\mu_{8}$
and $I_{20}$ calculated directly and via DTs, RF and a NN.}
\end{table}

The Table \ref{tab:=000020CPU-time=000020COS} illustrates the CPU
time of the COS method, where $L$ and $N$ are obtained by different
error tolerances. The COS method is implemented in C++ using for-loops
without parallelization. It is well known, that $L(\varepsilon,\mu_{8})$
is usually closer to the optimal truncation range than $L(\varepsilon,\mu_{4})$,
see \citet{junike2022precise}. It is therefore not surprising that
the average CPU time is about 10 times faster using the truncation
range $L(\varepsilon,\mu_{8})$ compared to $L(\varepsilon,\mu_{4})$,
see Table \ref{tab:=000020CPU-time=000020COS}.

\begin{table}[H]
\begin{centering}
\begin{tabular}{|c|c|c|c|c|c|c|}
\hline 
$\varepsilon$ & $\varepsilon=10^{-2}$ & $\varepsilon=10^{-3}$ & $\varepsilon=10^{-4}$ & $\varepsilon=10^{-5}$ & $\varepsilon=10^{-6}$ & $\varepsilon=10^{-7}$\tabularnewline
\hline 
\hline 
$L(\varepsilon,\mu_{4})$ & $5.89\cdot10^{-5}$ & $1.15\cdot10^{-4}$ & $2.34\cdot10^{-4}$ & $4.86\cdot10^{-4}$ & $1.02\cdot10^{-3}$ & $2.11\cdot10^{-3}$\tabularnewline
\hline 
$L(\varepsilon,\mu_{8})$ & $3.38\cdot10^{-5}$ & $4.66\cdot10^{-5}$ & $6.67\cdot10^{-5}$ & $9.80\cdot10^{-5}$ & $1.47\cdot10^{-4}$ & $2.20\cdot10^{-4}$\tabularnewline
\hline 
\end{tabular}
\par\end{centering}
\caption{\protect\label{tab:=000020CPU-time=000020COS}Average CPU time (in
sec.) on the test set of the COS method with truncation range $L(\varepsilon,\mu_{4})$
or $L(\varepsilon,\mu_{8})$ and number of terms $N(\varepsilon,I_{20})$
to price a call option with $S_{0}=100$ and $K=100$. Here, we only
take into account the CPU time of the COS method ignoring the CPU
time to estimate $L$ and $N$.}
\end{table}

\begin{table}[H]
\begin{centering}
\begin{tabular}{|>{\centering}p{2.5cm}|>{\centering}p{2.5cm}|>{\centering}p{2.5cm}|>{\centering}p{2.5cm}|>{\centering}p{2.5cm}|}
\hline 
Ana. calc. of $\mu_{4}$ and num. integration of $I_{20}$ & $\mu_{8}$ and $I_{20}$ via RF & $\mu_{8}$ and $I_{20}$ via NN & $\mu_{8}$ and $I_{20}$ via bDT & $\mu_{8}$ and $I_{20}$ via sDT\tabularnewline
\hline 
\hline 
$1.122\cdot10^{-2}$ & $6.921\cdot10^{-4}$ & $7.056\cdot10^{-5}$ & $2.607\cdot10^{-6}$ & $2.036\cdot10^{-6}$\tabularnewline
\hline 
\end{tabular}
\par\end{centering}
\caption{\protect\label{tab:=000020CPU-time=000020bound=000020moment}Average
CPU time on the test set in sec. for calculating $\mu_{n}$ and $I_{20}$
directly or using machine learning techniques.}
\end{table}

Let us set $\varepsilon=10^{-4}$ and let us consider two scenarios:
i) A trader estimates $\mu_{4}$ and $I_{20}$ directly. (Estimating
$\mu_{8}$ directly is too time consuming for the Heston model). ii)
A trader estimates $\mu_{8}$ and $I_{20}$ using machine learning
techniques. From Table \ref{tab:Accuracy=000020COS}, we can see that
both approaches will price the options very accurately for different
error tolerances and parameters of the Heston model. What is the impact
on the total CPU time? As shown in Table \ref{tab:=000020CPU-time=000020bound=000020moment},
the CPU time to obtain $\mu_{4}$ and $I_{20}$ directly takes about
0.011sec. (Most of the time is used to estimate $I_{20}$, we used
R's function \emph{integrate} with default values for numerical integration).
The computation of $\mu_{4}$ and $I_{20}$ dominates the total CPU
time, since the pure application of the COS method takes about $2.34\cdot10^{-4}$
sec., see Table \ref{tab:=000020CPU-time=000020COS}. On the other
hand, the CPU time to estimate $\mu_{8}$ and $I_{20}$ using machine
learning techniques is about a factor of $100$ to $1,000$ times
faster than the direct computation of $\mu_{4}$ and $I_{20}$. The
total CPU time of the COS method estimating $\mu_{8}$ and $I_{20}$
via a NN is about $1.4\cdot10^{-4}$ sec. In summary, approach ii)
is almost $100$ times faster than approach i).

\subsection{\protect\label{subsec:On-the-parameters}On the tuning parameters
of the Carr-Madan formula}

In order to apply the Carr-Madan formula, one must specify three parameters,
namely the damping factor $\alpha>0$, the truncation range $M$ and
the number of grid points $N$. In the following, we use a NN and
a RF to estimate these parameters. We set $M=1200$ and determine
optimal parameters $\alpha$ and $N$ for the entire training set,
such that $N$ is minimal to achieve an error bound of $10^{-7}$.
We then train a NN and a RF to learn these optimal parameters. Since
the estimate $\hat{N}$ of the NN and the RF sometimes significantly
underestimates the true $N$, we double the output of the NN and the
RF to improve the accuracy of the Carr-Madan formula. This step was
not necessary for the COS method, since the theoretical formulas for
the truncation range and number of terms are larger than the minimal
truncation range and number of terms. 

To measure the accuracy of the Carr-Madan formula, we price a call
option with $S_{0}=K=100$ and $r=0$, using the predicted values
for $\alpha$ and $N$ of the NN and the RF. We obtain the required
accuracy of $\varepsilon=10^{-7}$ for $90.55\%$ and $93.49\%$ of
the samples in the test set for the RF and the NN, respectively. 

To compare these results, we also use standard parameters of the Carr-Madan
formula: \citet{carr1999option} suggest the default values $M=1024$
and $N=2^{12}$ as a rule of thumb. The Carr-Madan formula is very
sensitive with respect to the damping factor, we choose $\alpha=1.95$.
For these default values, the accuracy of $10^{-7}$ is reached in
only $18.33\%$ of the samples in the test set (any other fixed $\alpha$
leads to an even lower proportion). Consequently, RFs and NNs are
a useful tool for improving the accuracy of the Carr-Madan formula,
since there is no single damping factor $\alpha$ and number of grid
points $N$ for all cases.

\section{Conclusion\protect\label{sec:Conclusion}}

In this paper, we proposed an indirect use of machine learning to
improve the efficiency and accuracy of the Carr-Madan formula and
the COS method for option pricing. \citet{junike2022precise} and
\citet{junike2023handle} provide explicit bounds on the truncation
range and the number of terms to apply the COS method. These bounds
ensure that the COS method prices a European option within a predefined
error tolerance. It is generally time-consuming to obtain these bounds
using classical numerical tools. In this paper, we instead estimate
these bounds using machine learning techniques such as RF, DT and
NN. We summarize the advantages:
\begin{itemize}
\item Compared to directly estimating the option prices using machine learning
techniques as in \citet{liu2019neural,liu2019pricing} and \citet{de2018machine},
our approach allows for full error control.
\item Compared to estimating the bounds using classical numerical methods,
our approach is much faster: about a factor $100$.
\item Compared to using a fast rule of thumb (as proposed in \citet{fang2009novel}
and \citet{carr1999option}) to estimate the tuning parameters of
the COS method or the Carr-Madan formula, our approach is much more
reliable. For the COS method, see \citet{junike2022precise} for examples
where a rule of thumb based on cumulants leads to serious mispricing.
For the Carr-Madan formula, see Section \ref{subsec:On-the-parameters}.
\end{itemize}
We tested RF, DT and NN to estimate the bounds to obtain the truncation
range and the number of terms to apply the COS method. Among these
techniques, the RF works best (accurate on 100\% of the test set).
The NN has a similar performance. But even a small DT gives very satisfactory
results (accurate on 98.2\% of the test set). Estimation of the tuning
parameters of the Carr-Madan formula by a RF or a NN works in about
90\% of all samples in a test set.

\appendix

\section{Appendix}

\subsection{\protect\label{subsec:=000020tree=000020bound}Decision tree of depth
$5$ to predict $I_{20}$}

\begin{longtable}[c]{|c|c|c|c|>{\centering}p{2.5cm}|>{\centering}p{2.5cm}|c|}
\hline 
nodeID & leaf node & variable & split value & left-child\\
 (if variable $\leq$ split value) & right-child \\
(if variable $>$ split value) & prediction\tabularnewline
\hline 
\endfirsthead
\hline 
nodeID & leaf node & variable & split value & left-child\\
 (if variable $\leq$ split value) & right-child \\
(if variable $>$ split value) & prediction\tabularnewline
\hline 
\endhead
\hline 
0 & No & $T$ & 0.186064 & 1 & 2 & NA\tabularnewline
\hline 
1 & No & $v_{0}$ & 0.236779 & 3 & 4 & NA\tabularnewline
\hline 
2 & No & $T$ & 1.143101 & 5 & 6 & NA\tabularnewline
\hline 
3 & No & $T$ & 0.062439 & 7 & 8 & NA\tabularnewline
\hline 
4 & No & $\xi$ & 2.705391 & 9 & 10 & NA\tabularnewline
\hline 
5 & No & $\xi$ & 2.436885 & 11 & 12 & NA\tabularnewline
\hline 
6 & No & $T$ & 2.887055 & 13 & 14 & NA\tabularnewline
\hline 
7 & No & $v_{0}$ & 0.022762 & 15 & 16 & NA\tabularnewline
\hline 
8 & No & $\rho$ & 0.976444 & 17 & 18 & NA\tabularnewline
\hline 
9 & No & $T$ & 0.020387 & 19 & 20 & NA\tabularnewline
\hline 
10 & No & $v_{0}$ & 0.698183 & 21 & 22 & NA\tabularnewline
\hline 
11 & No & $T$ & 0.420034 & 23 & 24 & NA\tabularnewline
\hline 
12 & No & $T$ & 0.527950 & 25 & 26 & NA\tabularnewline
\hline 
13 & No & $\theta$ & 0.784247 & 27 & 28 & NA\tabularnewline
\hline 
14 & No & $\theta$ & 0.640466 & 29 & 30 & NA\tabularnewline
\hline 
15 & No & $\rho$ & -0.468963 & 31 & 32 & NA\tabularnewline
\hline 
16 & No & $\xi$ & 2.258602 & 33 & 34 & NA\tabularnewline
\hline 
17 & No & $\xi$ & 2.470854 & 35 & 36 & NA\tabularnewline
\hline 
18 & Yes & NA & NA & NA & NA & 429.628317\tabularnewline
\hline 
19 & No & $v_{0}$ & 0.587902 & 37 & 38 & NA\tabularnewline
\hline 
20 & No & $v_{0}$ & 0.694761 & 39 & 40 & NA\tabularnewline
\hline 
21 & No & $\rho$ & 0.806959 & 41 & 42 & NA\tabularnewline
\hline 
22 & No & $\rho$ & -0.965657 & 43 & 44 & NA\tabularnewline
\hline 
23 & No & $\rho$ & 0.960696 & 45 & 46 & NA\tabularnewline
\hline 
24 & No & $\theta$ & 0.719155 & 47 & 48 & NA\tabularnewline
\hline 
25 & No & $\rho$ & 0.910680 & 49 & 50 & NA\tabularnewline
\hline 
26 & No & $\rho$ & 0.971400 & 51 & 52 & NA\tabularnewline
\hline 
27 & No & $\xi$ & 1.991873 & 53 & 54 & NA\tabularnewline
\hline 
28 & No & $\kappa$ & 3.484651 & 55 & 56 & NA\tabularnewline
\hline 
29 & No & $T$ & 5.496469 & 57 & 58 & NA\tabularnewline
\hline 
30 & No & $T$ & 5.071144 & 59 & 60 & NA\tabularnewline
\hline 
31 & Yes & NA & NA & NA & NA & 487.342705\tabularnewline
\hline 
32 & Yes & NA & NA & NA & NA & 235.195790\tabularnewline
\hline 
33 & Yes & NA & NA & NA & NA & 102.893171\tabularnewline
\hline 
34 & Yes & NA & NA & NA & NA & 201.553675\tabularnewline
\hline 
35 & Yes & NA & NA & NA & NA & 36.203604\tabularnewline
\hline 
36 & Yes & NA & NA & NA & NA & 88.277112\tabularnewline
\hline 
37 & Yes & NA & NA & NA & NA & 62.208418\tabularnewline
\hline 
38 & Yes & NA & NA & NA & NA & 32.587266\tabularnewline
\hline 
39 & Yes & NA & NA & NA & NA & 25.738887\tabularnewline
\hline 
40 & Yes & NA & NA & NA & NA & 14.188984\tabularnewline
\hline 
41 & Yes & NA & NA & NA & NA & 64.373567\tabularnewline
\hline 
42 & Yes & NA & NA & NA & NA & 145.963858\tabularnewline
\hline 
43 & Yes & NA & NA & NA & NA & 133.842206\tabularnewline
\hline 
44 & Yes & NA & NA & NA & NA & 31.212500\tabularnewline
\hline 
45 & Yes & NA & NA & NA & NA & 9.945991\tabularnewline
\hline 
46 & Yes & NA & NA & NA & NA & 38.703759\tabularnewline
\hline 
47 & Yes & NA & NA & NA & NA & 8.617465\tabularnewline
\hline 
48 & Yes & NA & NA & NA & NA & 4.853007\tabularnewline
\hline 
49 & Yes & NA & NA & NA & NA & 18.833877\tabularnewline
\hline 
50 & Yes & NA & NA & NA & NA & 47.724354\tabularnewline
\hline 
51 & Yes & NA & NA & NA & NA & 10.171048\tabularnewline
\hline 
52 & Yes & NA & NA & NA & NA & 45.038898\tabularnewline
\hline 
53 & Yes & NA & NA & NA & NA & 4.335898\tabularnewline
\hline 
54 & Yes & NA & NA & NA & NA & 7.306731\tabularnewline
\hline 
55 & Yes & NA & NA & NA & NA & 4.961622\tabularnewline
\hline 
56 & Yes & NA & NA & NA & NA & 2.733986\tabularnewline
\hline 
57 & Yes & NA & NA & NA & NA & 3.421163\tabularnewline
\hline 
58 & Yes & NA & NA & NA & NA & 2.172563\tabularnewline
\hline 
59 & Yes & NA & NA & NA & NA & 1.894587\tabularnewline
\hline 
60 & Yes & NA & NA & NA & NA & 1.144569\tabularnewline
\hline 
\end{longtable}

\subsection{\protect\label{subsec:=000020tree=000020m8}Decision tree of depth
$5$ to predict $\mu_{8}$}

\begin{longtable}[c]{|c|c|c|c|>{\centering}p{2.5cm}|>{\centering}p{2.5cm}|c|}
\hline 
nodeID & leaf node & variable & splitvalue & left-child\\
 (if variable $\leq$ splitvalue) & right-child \\
(if variable $>$ splitvalue) & prediction\tabularnewline
\hline 
\endfirsthead
\hline 
nodeID & leaf node & variable & splitvalue & left-child\\
 (if variable $\leq$ splitvalue) & right-child \\
(if variable $>$ splitvalue) & prediction\tabularnewline
\hline 
\endhead
\hline 
0 & No & $T$ & 3.399204 & 1 & 2 & NA\tabularnewline
\hline 
1 & No & $T$ & 1.169626 & 3 & 4 & NA\tabularnewline
\hline 
2 & No & $\theta$ & 0.959098 & 5 & 6 & NA\tabularnewline
\hline 
3 & No & $T$ & 0.428831 & 7 & 8 & NA\tabularnewline
\hline 
4 & No & $\theta$ & 0.900109 & 9 & 10 & NA\tabularnewline
\hline 
5 & No & $\theta$ & 0.498129 & 11 & 12 & NA\tabularnewline
\hline 
6 & No & $\kappa$ & 2.697343 & 13 & 14 & NA\tabularnewline
\hline 
7 & No & $T$ & 0.183570 & 15 & 16 & NA\tabularnewline
\hline 
8 & No & $\xi$ & 2.347370 & 17 & 18 & NA\tabularnewline
\hline 
9 & No & $\theta$ & 0.417688 & 19 & 20 & NA\tabularnewline
\hline 
10 & No & $\rho$ & -0.102140 & 21 & 22 & NA\tabularnewline
\hline 
11 & No & $\theta$ & 0.273765 & 23 & 24 & NA\tabularnewline
\hline 
12 & No & $\rho$ & -0.192910 & 25 & 26 & NA\tabularnewline
\hline 
13 & No & $\xi$ & 2.690779 & 27 & 28 & NA\tabularnewline
\hline 
14 & No & $\rho$ & -0.116578 & 29 & 30 & NA\tabularnewline
\hline 
15 & No & $T$ & 0.074047 & 31 & 32 & NA\tabularnewline
\hline 
16 & No & $\xi$ & 2.265810 & 33 & 34 & NA\tabularnewline
\hline 
17 & No & $\theta$ & 0.834463 & 35 & 36 & NA\tabularnewline
\hline 
18 & No & $\rho$ & -0.186050 & 37 & 38 & NA\tabularnewline
\hline 
19 & No & $\theta$ & 0.226219 & 39 & 40 & NA\tabularnewline
\hline 
20 & No & $\rho$ & -0.354863 & 41 & 42 & NA\tabularnewline
\hline 
21 & No & $\xi$ & 2.903877 & 43 & 44 & NA\tabularnewline
\hline 
22 & No & $T$ & 2.237251 & 45 & 46 & NA\tabularnewline
\hline 
23 & No & $\theta$ & 0.177632 & 47 & 48 & NA\tabularnewline
\hline 
24 & No & $\rho$ & -0.286902 & 49 & 50 & NA\tabularnewline
\hline 
25 & No & $\xi$ & 2.631217 & 51 & 52 & NA\tabularnewline
\hline 
26 & No & $T$ & 6.095978 & 53 & 54 & NA\tabularnewline
\hline 
27 & No & $\rho$ & -0.015948 & 55 & 56 & NA\tabularnewline
\hline 
28 & No & $\rho$ & -0.210373 & 57 & 58 & NA\tabularnewline
\hline 
29 & No & $\xi$ & 3.133527 & 59 & 60 & NA\tabularnewline
\hline 
30 & No & $T$ & 6.186172 & 61 & 62 & NA\tabularnewline
\hline 
31 & Yes & NA & NA & NA & NA & 0.366800\tabularnewline
\hline 
32 & Yes & NA & NA & NA & NA & 0.754696\tabularnewline
\hline 
33 & Yes & NA & NA & NA & NA & 1.068672\tabularnewline
\hline 
34 & Yes & NA & NA & NA & NA & 1.464186\tabularnewline
\hline 
35 & Yes & NA & NA & NA & NA & 1.367173\tabularnewline
\hline 
36 & Yes & NA & NA & NA & NA & 1.973656\tabularnewline
\hline 
37 & Yes & NA & NA & NA & NA & 3.110276\tabularnewline
\hline 
38 & Yes & NA & NA & NA & NA & 2.107476\tabularnewline
\hline 
39 & Yes & NA & NA & NA & NA & 1.381474\tabularnewline
\hline 
40 & Yes & NA & NA & NA & NA & 2.026955\tabularnewline
\hline 
41 & Yes & NA & NA & NA & NA & 3.456599\tabularnewline
\hline 
42 & Yes & NA & NA & NA & NA & 2.541016\tabularnewline
\hline 
43 & Yes & NA & NA & NA & NA & 3.922665\tabularnewline
\hline 
44 & Yes & NA & NA & NA & NA & 5.965712\tabularnewline
\hline 
45 & Yes & NA & NA & NA & NA & 2.993397\tabularnewline
\hline 
46 & Yes & NA & NA & NA & NA & 3.799591\tabularnewline
\hline 
47 & Yes & NA & NA & NA & NA & 1.785174\tabularnewline
\hline 
48 & Yes & NA & NA & NA & NA & 2.496068\tabularnewline
\hline 
49 & Yes & NA & NA & NA & NA & 3.837462\tabularnewline
\hline 
50 & Yes & NA & NA & NA & NA & 3.023879\tabularnewline
\hline 
51 & Yes & NA & NA & NA & NA & 4.734888\tabularnewline
\hline 
52 & Yes & NA & NA & NA & NA & 6.274068\tabularnewline
\hline 
53 & Yes & NA & NA & NA & NA & 3.484621\tabularnewline
\hline 
54 & Yes & NA & NA & NA & NA & 4.394604\tabularnewline
\hline 
55 & Yes & NA & NA & NA & NA & 8.766430\tabularnewline
\hline 
56 & Yes & NA & NA & NA & NA & 5.524519\tabularnewline
\hline 
57 & Yes & NA & NA & NA & NA & 17.084012\tabularnewline
\hline 
58 & Yes & NA & NA & NA & NA & 10.369077\tabularnewline
\hline 
59 & Yes & NA & NA & NA & NA & 6.240159\tabularnewline
\hline 
60 & Yes & NA & NA & NA & NA & 8.173560\tabularnewline
\hline 
61 & Yes & NA & NA & NA & NA & 4.647802\tabularnewline
\hline 
62 & Yes & NA & NA & NA & NA & 5.944725\tabularnewline
\hline 
\end{longtable}

\subsection{\protect\label{subsec:=000020code}Simple implementation }

The following algorithm implements the COS method in R for the Heston
model to price European put and call options. 

\rule[0.5ex]{1\columnwidth}{1pt}

\textbf{Algorithm 1} Implementation details of the COS method in the
Heston model

\rule[0.5ex]{1\columnwidth}{0.5pt}

{\footnotesize\#Characteristic function of log-returns in the Heston
with parameters params.}{\footnotesize\par}

{\footnotesize\#The characteristic function is taken from Schoutens
et. al (2004).}{\footnotesize\par}

{\footnotesize psiLogST\_Heston = function(u, mat, params, S0, r)\{}{\footnotesize\par}

{\footnotesize ~~~kappa = params{[}1{]} \#speed of mean reversion}{\footnotesize\par}

{\footnotesize ~~~theta = params{[}2{]} \#level of mean reversion}{\footnotesize\par}

{\footnotesize ~~~xi = params{[}3{]} \#vol of vol}{\footnotesize\par}

{\footnotesize ~~~rho = params{[}4{]} \#correlation vol stock}{\footnotesize\par}

{\footnotesize ~~~v0 = params{[}5{]} \#initial vol}{\footnotesize\par}

{\footnotesize ~~~d = sqrt((rho {*} xi {*} u {*} 1i - kappa)\textasciicircum 2
- xi\textasciicircum 2 {*} (-1i {*} u - u\textasciicircum 2))}{\footnotesize\par}

{\footnotesize ~~~mytmp = kappa - rho {*} xi {*} u {*} 1i}{\footnotesize\par}

{\footnotesize ~~~g = (mytmp - d) / (mytmp + d)}{\footnotesize\par}

{\footnotesize ~~~expdmat = exp(-d {*} mat)}{\footnotesize\par}

{\footnotesize ~~~tmp0 = 1i {*} u {*} (log(S0) + r {*} mat)}{\footnotesize\par}

{\footnotesize ~~~tmp1 = (mytmp - d) {*} mat - 2 {*} log((1 - g
{*} expdmat) / (1 - g))}{\footnotesize\par}

{\footnotesize ~~~tmp2 = theta {*} kappa {*} xi\textasciicircum (-2)
{*} tmp1}{\footnotesize\par}

{\footnotesize ~~~tmp3 = v0 {*} xi\textasciicircum (-2) {*} (mytmp
- d) {*} (1 - expdmat) / (1 - g {*} expdmat)}{\footnotesize\par}

{\footnotesize ~~~exp(tmp0 + tmp2 + tmp3)}{\footnotesize\par}

{\footnotesize\}}{\footnotesize\par}

{\footnotesize library(Deriv) \#There are much faster alternatives
like SageMath.}{\footnotesize\par}

{\footnotesize psiLogST\_Heston1=Deriv(psiLogST\_Heston, \textquotedbl u\textquotedbl ) }{\footnotesize\par}

{\footnotesize ~~~}{\footnotesize\par}

{\footnotesize\#mu is equal to E{[}log(S\_T){]}}{\footnotesize\par}

{\footnotesize mu = function(mat, params, S0, r)\{}{\footnotesize\par}

{\footnotesize ~~~Re(-1i {*} psiLogST\_Heston1(0, mat, params,
S0, r))}{\footnotesize\par}

{\footnotesize\}}{\footnotesize\par}

{\footnotesize\#Characteristic function of centralized log-returns
in the Heston model.}{\footnotesize\par}

{\footnotesize phi = function(u, mat, params, S0, r)\{}{\footnotesize\par}

{\footnotesize ~~~psiLogST\_Heston(u, mat, params, S0, r) {*} exp(-1i
{*} u {*} mu(mat, params, S0, r))}{\footnotesize\par}

{\footnotesize\}}{\footnotesize\par}

{\footnotesize\#cosine coefficients of the density.}{\footnotesize\par}

{\footnotesize ck = function(L, mat, N, params, S0, r)\{}{\footnotesize\par}

{\footnotesize ~~~k = 0:N}{\footnotesize\par}

{\footnotesize ~~~return(1 / L {*} Re(phi(k {*} pi / (2 {*} L),
mat, params, S0, r) {*}  exp(1i {*} k {*} pi/2)))}{\footnotesize\par}

{\footnotesize\}}{\footnotesize\par}

{\footnotesize\#cosine coefficients of a put option, see Appendix
Junike and Pankrashkin (2022).}{\footnotesize\par}

{\footnotesize vk = function(K, L, mat, N, params, S0, r)\{}{\footnotesize\par}

{\footnotesize ~~~mymu = mu(mat, params, S0, r)   \#mu = E{[}log(S\_T){]}}{\footnotesize\par}

{\footnotesize ~~~d = min(log(K) - mymu, L)}{\footnotesize\par}

{\footnotesize ~~~if(d <= -L)}{\footnotesize\par}

{\footnotesize ~~~~~~return(rep(0, N + 1)) \#Return zero vector}{\footnotesize\par}

{\footnotesize ~~~k = 0:N }{\footnotesize\par}

{\footnotesize ~~~psi0 = 2 {*} L / (k {*} pi) {*} (sin(k {*} pi
{*} (d + L) / (2 {*} L)))}{\footnotesize\par}

{\footnotesize ~~~psi0{[}1{]} = d + L}{\footnotesize\par}

{\footnotesize ~~~tmp1 = k {*} pi / (2 {*} L) {*} sin( k {*} pi
{*} (d + L) / (2 {*} L))}{\footnotesize\par}

{\footnotesize ~~~tmp2 = cos(k {*} pi {*} (d + L) / (2 {*} L))}{\footnotesize\par}

{\footnotesize ~~~tmp3 = 1 + (k {*} pi / (2 {*} L))\textasciicircum 2}{\footnotesize\par}

{\footnotesize ~~~psi1 = (exp(d) {*} (tmp1 + tmp2) - exp(-L)) /
tmp3}{\footnotesize\par}

{\footnotesize ~~~return(exp(-r {*} mat) {*} (K {*} psi0 - exp(mymu)
{*} psi1))}{\footnotesize\par}

{\footnotesize\}}{\footnotesize\par}

{\footnotesize\#approximation of put option by COS method}{\footnotesize\par}

{\footnotesize put\_COS = function(K, L, mat, N, params, S0, r)\{}{\footnotesize\par}

{\footnotesize ~~~tmp = ck(L, mat, N, params, S0, r) {*} vk(K,
L, mat, N, params, S0, r)}{\footnotesize\par}

{\footnotesize ~~~tmp{[}1{]} = 0.5 {*} tmp{[}1{]} \#First term
is weighted by 1/2}{\footnotesize\par}

{\footnotesize ~~~return(sum(tmp))}{\footnotesize\par}

{\footnotesize\}}{\footnotesize\par}

{\footnotesize ~~~}{\footnotesize\par}

{\footnotesize\#approximation of call option by COS method using put-call
parity}{\footnotesize\par}

{\footnotesize call\_COS = function(K, L, mat, N, params, S0, r)\{}{\footnotesize\par}

{\footnotesize ~~~return(put\_COS(K, L, mat, N, params, S0, r)
+ S0 - K {*} exp(-r {*} mat))}{\footnotesize\par}

{\footnotesize\}}{\footnotesize\par}

{\footnotesize ~~~}{\footnotesize\par}

{\footnotesize\#Derivatives of the characteristic function of the
centralized log-returns in the Heston model.}{\footnotesize\par}

{\footnotesize phi1 = Deriv(phi, \textquotedbl u\textquotedbl )}{\footnotesize\par}

{\footnotesize phi2 = Deriv(phi1, \textquotedbl u\textquotedbl )}{\footnotesize\par}

{\footnotesize phi3 = Deriv(phi2, \textquotedbl u\textquotedbl )
\#Takes very long but has to be done only once.}{\footnotesize\par}

{\footnotesize phi4 = Deriv(phi3, \textquotedbl u\textquotedbl )
\#Takes very long but has to be done only once.}{\footnotesize\par}

{\footnotesize save(phi4, file = \textquotedbl phi4.RData\textquotedbl )
\#save for later use. Load with load(\textquotedbl phi4.RData\textquotedbl ).}{\footnotesize\par}

{\footnotesize ~~~}{\footnotesize\par}

{\footnotesize\#Price a put option in the Heston model by the COS
method.}{\footnotesize\par}

{\footnotesize eps = 10\textasciicircum -6 \#error tolerance}{\footnotesize\par}

{\footnotesize K = 90 \#strike}{\footnotesize\par}

{\footnotesize S0 = 100 \#current stock price}{\footnotesize\par}

{\footnotesize r = 0.1 \#interest rates}{\footnotesize\par}

{\footnotesize params = c(0.6067, 0.0707, 0.2928, -0.7571, 0.0654)}{\footnotesize\par}

{\footnotesize mat = 0.7 \#maturity}{\footnotesize\par}

{\footnotesize mu\_n = abs(phi4(0, mat, params, S0, r)) \#4-th moment
of log-returns. }{\footnotesize\par}

{\footnotesize L = (2 {*} K {*} exp(-r {*} mat) {*} mu\_n / eps)\textasciicircum (1
/ 4) \#Junike (2024, Eq. (3.10)).}{\footnotesize\par}

{\footnotesize s = 20 \#number of derivatives to determine the number
of terms}{\footnotesize\par}

{\footnotesize integrand = function(u)\{1 / (2 {*} pi) {*} abs(u)\textasciicircum (s
+ 1) {*} abs(phi(u, mat, params, S0, r))\} }{\footnotesize\par}

{\footnotesize boundDeriv = integrate(integrand, -Inf, Inf)\$value}{\footnotesize\par}

{\footnotesize tmp = 2\textasciicircum (s + 5 / 2) {*} boundDeriv
{*} L\textasciicircum (s + 2) {*} 12 {*} K {*} exp(-r {*} mat)}{\footnotesize\par}

{\footnotesize N = ceiling((tmp / (s {*} pi\textasciicircum (s + 1)
{*} eps))\textasciicircum (1 / s)) \#Number of terms, Junike (2024,
Sec. 6.1)   }{\footnotesize\par}

{\footnotesize put\_COS(K, L, mat, N, params, S0, r) \#The price of
put option is 2.773954.}{\footnotesize\par}

\rule[0.5ex]{1\columnwidth}{0.5pt}

\bibliographystyle{plainnat}
\bibliography{biblio}

\begin{thebibliography}{26}
\providecommand{\natexlab}[1]{#1}
\providecommand{\url}[1]{\texttt{#1}}
\expandafter\ifx\csname urlstyle\endcsname\relax
  \providecommand{\doi}[1]{doi: #1}\else
  \providecommand{\doi}{doi: \begingroup \urlstyle{rm}\Url}\fi

\bibitem[Andersen(2008)]{andersen2007efficient}
L.~B.~G. Andersen.
\newblock {Efficient simulation of the Heston stochastic volatility model}.
\newblock \emph{Journal of Computational Finance}, 11\penalty0 (3):\penalty0
  1--43, 2008.

\bibitem[Andersen and Piterbarg(2007)]{andersen2007moment}
L.~B.~G. Andersen and V.~V. Piterbarg.
\newblock Moment explosions in stochastic volatility models.
\newblock \emph{Finance and Stochastics}, 11\penalty0 (1):\penalty0 29--50,
  2007.

\bibitem[Bakshi et~al.(1997)Bakshi, Cao, and Chen]{bakshi1997empirical}
G.~Bakshi, C.~Cao, and Z.~Chen.
\newblock Empirical performance of alternative option pricing models.
\newblock \emph{The Journal of Finance}, 52\penalty0 (5):\penalty0 2003--2049,
  1997.

\bibitem[Breiman(2001)]{breiman2001random}
L.~Breiman.
\newblock Random forests.
\newblock \emph{Machine Learning}, 45:\penalty0 5--32, 2001.

\bibitem[Breiman et~al.(1984)Breiman, Friedman, Olshen, and
  Stone]{breiman1984classification}
L.~Breiman, J.~Friedman, R.~Olshen, and C.~Stone.
\newblock \emph{Classification and regression trees}.
\newblock Taylor \& Francis, 1984.

\bibitem[Carr and Madan(1999)]{carr1999option}
P.~Carr and D.~Madan.
\newblock {Option valuation using the fast Fourier transform}.
\newblock \emph{Journal of Computational Finance}, 2\penalty0 (4):\penalty0
  61--73, 1999.

\bibitem[Choudhury and Lucantoni(1996)]{choudhury1996numerical}
G.~L. Choudhury and D.~M. Lucantoni.
\newblock Numerical computation of the moments of a probability distribution
  from its transform.
\newblock \emph{Operations Research}, 44\penalty0 (2):\penalty0 368--381, 1996.

\bibitem[Cris{\'o}stomo(2015)]{crisostomo2015analysis}
R.~Cris{\'o}stomo.
\newblock {An analysis of the Heston stochastic volatility model:
  Implementation and calibration using matlab}.
\newblock \emph{arXiv preprint arXiv:1502.02963}, 2015.

\bibitem[Cui et~al.(2017)Cui, del Ba{\~n}o~Rollin, and Germano]{cui2017full}
Y.~Cui, S.~del Ba{\~n}o~Rollin, and G.~Germano.
\newblock {Full and fast calibration of the Heston stochastic volatility
  model}.
\newblock \emph{European Journal of Operational Research}, 263\penalty0
  (2):\penalty0 625--638, 2017.

\bibitem[De~Spiegeleer et~al.(2018)De~Spiegeleer, Madan, Reyners, and
  Schoutens]{de2018machine}
J.~De~Spiegeleer, D.~B. Madan, S.~Reyners, and W.~Schoutens.
\newblock Machine learning for quantitative finance: fast derivative pricing,
  hedging and fitting.
\newblock \emph{Quantitative Finance}, 18\penalty0 (10):\penalty0 1635--1643,
  2018.

\bibitem[Engelmann et~al.(2021)Engelmann, Koster, and
  Oeltz]{engelmann2021calibration}
B.~Engelmann, F.~Koster, and D.~Oeltz.
\newblock {Calibration of the Heston stochastic local volatility model: A
  finite volume scheme}.
\newblock \emph{International Journal of Financial Engineering}, 8\penalty0
  (01):\penalty0 2050048, 2021.

\bibitem[Fang and Oosterlee(2009)]{fang2009novel}
F.~Fang and C.~W. Oosterlee.
\newblock {A novel pricing method for European options based on Fourier-cosine
  series expansions}.
\newblock \emph{SIAM Journal on Scientific Computing}, 31\penalty0
  (2):\penalty0 826--848, 2009.

\bibitem[Forde et~al.(2012)Forde, Jacquier, and Lee]{forde2012small}
M.~Forde, A.~Jacquier, and R.~Lee.
\newblock {The small-time smile and term structure of implied volatility under
  the Heston model}.
\newblock \emph{SIAM Journal on Financial Mathematics}, 3\penalty0
  (1):\penalty0 690--708, 2012.

\bibitem[Goodfellow et~al.(2016)Goodfellow, Bengio, and
  Courville]{goodfellow2016deep}
I.~Goodfellow, Y.~Bengio, and A.~Courville.
\newblock \emph{Deep Learning}.
\newblock MIT Press, 2016.
\newblock \url{http://www.deeplearningbook.org}.

\bibitem[He et~al.(2015)He, Zhang, Ren, and Sun]{he2015delving}
Kaiming He, Xiangyu Zhang, Shaoqing Ren, and Jian Sun.
\newblock Delving deep into rectifiers: Surpassing human-level performance on
  imagenet classification.
\newblock In \emph{Proceedings of the IEEE international conference on computer
  vision}, pages 1026--1034, 2015.

\bibitem[Heston(1993)]{heston1993closed}
S.~L. Heston.
\newblock {A closed-form solution for options with stochastic volatility with
  applications to bond and currency options}.
\newblock \emph{The Review of Financial Studies}, 6\penalty0 (2):\penalty0
  327--343, 1993.

\bibitem[Junike(2024)]{junike2023handle}
G.~Junike.
\newblock {On the number of terms in the COS method for European option
  pricing}.
\newblock \emph{Numerische Mathematik}, 156\penalty0 (2):\penalty0 533--564,
  2024.

\bibitem[Junike and Pankrashkin(2022)]{junike2022precise}
G.~Junike and K.~Pankrashkin.
\newblock {Precise option pricing by the COS method--How to choose the
  truncation range}.
\newblock \emph{Applied Mathematics and Computation}, 421:\penalty0 126935,
  2022.

\bibitem[Levendorski{\u{i}}(2012)]{levendorskiui2012efficient}
S.~Levendorski{\u{i}}.
\newblock {Efficient pricing and reliable calibration in the Heston model}.
\newblock \emph{International Journal of Theoretical and Applied Finance},
  15\penalty0 (07):\penalty0 1250050, 2012.

\bibitem[Liu et~al.(2019{\natexlab{a}})Liu, Borovykh, Grzelak, and
  Oosterlee]{liu2019neural}
S.~Liu, A.~Borovykh, L.~A. Grzelak, and C.~W. Oosterlee.
\newblock {A neural network-based framework for financial model calibration}.
\newblock \emph{Journal of Mathematics in Industry}, 9\penalty0 (1):\penalty0
  1--28, 2019{\natexlab{a}}.

\bibitem[Liu et~al.(2019{\natexlab{b}})Liu, Oosterlee, and
  Bohte]{liu2019pricing}
S.~Liu, C.~W. Oosterlee, and S.~M. Bohte.
\newblock {Pricing options and computing implied volatilities using neural
  networks}.
\newblock \emph{Risks}, 7\penalty0 (1):\penalty0 1--22, 2019{\natexlab{b}}.

\bibitem[Oosterlee and Grzelak(2019)]{oosterlee2019mathematical}
C.~W. Oosterlee and L.~A. Grzelak.
\newblock \emph{{Mathematical modeling and computation in finance: with
  exercises and Python and MATLAB computer codes}}.
\newblock World Scientific, 2019.

\bibitem[Ruf and Wang(2020)]{ruf2019neural}
Johannes Ruf and Weiguan Wang.
\newblock Neural networks for option pricing and hedging: a literature review.
\newblock \emph{Journal of Computational Finance}, 2020.

\bibitem[Schoutens et~al.(2003)Schoutens, Simons, and
  Tistaert]{schoutens2003perfect}
W.~Schoutens, E.~Simons, and J.~Tistaert.
\newblock {A perfect calibration! Now what?}
\newblock \emph{The best of Wilmott}, pages 281--304, 2003.

\bibitem[Sirignano and Spiliopoulos(2018)]{sirignano2018dgm}
J.~Sirignano and K.~Spiliopoulos.
\newblock {DGM: A deep learning algorithm for solving partial differential
  equations}.
\newblock \emph{Journal of Computational Physics}, 375:\penalty0 1339--1364,
  2018.

\bibitem[Yang et~al.(2017)Yang, Zheng, and Hospedales]{yang2017gated}
Y.~Yang, Y.~Zheng, and T.~Hospedales.
\newblock {Gated neural networks for option pricing: Rationality by design}.
\newblock In \emph{Proceedings of the AAAI conference on artificial
  intelligence}, volume~31, 2017.

\end{thebibliography}

\end{document}